\documentclass[letterpaper,9pt,conference]{IEEEtran}
\usepackage{graphicx,latexsym,epsfig,amssymb,amsmath,subfigure}

\begin{document}

\title{Cyber-Insurance in Internet Security \\ \emph{A Dig into the Information Asymmetry Problem}}



\author{\IEEEauthorblockN{Ranjan Pal}
\IEEEauthorblockA{Department of Computer Science\\
University of Southern California\\
Email: rpal@usc.edu}}


%

\small
\maketitle
\begin{abstract}
Internet users such as individuals and organizations are subject to different types of epidemic risks such as worms, viruses, spams, and botnets. To reduce the probability of risk, an Internet user generally invests in traditional security mechanisms like anti-virus and anti-spam software, sometimes also known as \emph{self-defense} mechanisms. However, according to security experts, such software (and their subsequent advancements) will not completely eliminate risk. Recent research efforts have considered the problem of residual risk elimination by proposing the idea of \emph{cyber-insurance}. In this regard, an important research problem is resolving information asymmetry issues associated with cyber-insurance contracts. In this paper we propose \emph{three} mechanisms to resolve information asymmetry in cyber-insurance. Our mechanisms are based on the \emph{Principal-Agent} (PA) model in microeconomic theory.
We show that (1) optimal cyber-insurance contracts induced by our mechanisms only provide partial coverage to the insureds. This ensures greater self-defense efforts on the part of the latter to protect their computing systems, which in turn increases overall network security, (2) the level of deductible per network user contract increases in a concave manner with the topological degree of the user, and (3) a market for cyber-insurance can be made to exist in the presence of monopolistic insurers under effective mechanism design. Our methodology is applicable to any distributed network scenario in which a framework for cyber-insurance can be implemented. 

\end{abstract}
%
\IEEEpeerreviewmaketitle
\section{Introduction}
The Internet has become a fundamental and an integral part of our daily lives. Billions of people nowadays are using the Internet for various types of applications. However, all these applications are running on a network, that was built under assumptions, some of which are no longer valid for today's applications, e,g., that all users on the Internet can be trusted and that there are no malicious elements propagating in the Internet. On the contrary, the infrastructure, the users, and the services offered on the Internet today are all subject to a wide variety of risks. These risks include distributed denial of service attacks, intrusions of various kinds, hacking, phishing, worms, viruses, spams, etc. In order to counter the threats posed by the risks, Internet users\footnote{The term `users' may refer to both, individuals and organizations. The network under consideration may be the Internet or any other distributed communication network where users have access to the Internet.} have traditionally resorted to antivirus and anti-spam softwares, firewalls, and other add-ons to reduce the likelihood of being affected by threats. In practice, a large industry (companies like \emph{Symantec, McAfee,} etc.) as well as considerable research efforts are currently centered around developing and deploying tools and techniques to detect threats and anomalies in order to protect the Internet infrastructure and its users from the negative impact of the anomalies. However, security experts \cite{amr} claim that it is impossible to achieve perfect/near-perfect Internet security just via technological advancements. 

\subsection{Why Technological Advancements Aren't Enough?}
In the past one and half decade, risk protection techniques from a variety of computer science fields such as cryptography, hardware engineering, and software engineering have continually made improvements. Inspite of such improvements, recent articles by Anderson \cite{ranr}\cite{amr}\cite{ramr} have stated that it is impossible to achieve a 100\% Internet security protection. The authors attribute this impossibility primarily to six reasons: 
\begin{enumerate}
\item Existing technical solutions are not sound, i.e.,there do not always exist fool-proof ways to detect and identify even well  deÞned threats; for example, even state of the art detectors of port scanners and other known anomalies suffer from positive rates of false positives and false negatives \cite{jpbb}. In addition, the originators of  threats, and the threats they produce, evolve automatically in response to detection and mitigation solutions being deployed, which makes it harder to detect and mitigate evolving threat signatures and characteristics \cite{vg}. Finally, completely eliminating risks would require the use of formal methods to design provably secure systems - however, these methods capture with difficulty the presence of those messy humans, even non malicious humans, in the loop \cite{odlyzko}. 
\item The Internet is a distributed system, where the system users have divergent security interests and incentives, leading to the problem of `misaligned incentives' amongst users. For example, a rational Internet user might well spend \$20 to stop a virus trashing its hard disk, but would hardly have any incentive to invest sufficient amounts in security solutions to prevent its computer being used by an attacker for a service-denial attack on a wealthy corporation like an Amazon or a Microsoft \cite{varian}. Thus, it is evident that the problem of misaligned incentives can be resolved only if liabilities are assigned to parties (users) that can best manage risk. 
\item The risks faced by Internet users are often correlated and interdependent. As a result a user taking protective action in an Internet like distributed system creates positive externalities \cite{hh} for other networked users that in turn may discourage them from making appropriate security investments, leading to the `free-riding' problem \cite{gccr}\cite{jaw}\cite{mybm}\cite{oom}. The free-riding problem leads to suboptimal network security. 
\item Network externalities due to \emph{lock-in} and \emph{first-mover} effects of security software vendors affect the adoption of more advanced technology \cite{ranr}.
\item Many security software markets have aspects of a \emph{lemons market} \cite{lmr} or even worse, i.e., by looking at security software, even the vendor does not know how secure its software is \cite{ramr}. So buyers have no reason to pay for more protection, and vendors are disinclined to invest time, money, and effort to strengthen their security software code. 
\end{enumerate}

\subsection{The Advent of Cyber-insurance}
In view of the above mentioned inevitable barriers to 100\% risk mitigation, the need arises for alternative methods of risk management in the Internet. Anderson and Moore \cite{amr} state that \emph{microeconomics}, \emph{game theory}, and \emph{psychology} will play as vital a role in effective risk management in the modern and future Internet, as did the mathematics of cryptography a quarter century ago. In this regard, \emph{cyber-insurance} is a psycho-economic-driven risk-management technique, where risks are transferred to a third party, i.e., an insurance company, in return for a fee, i.e., the \emph{insurance premium}. The concept of cyber-insurance is growing in importance amongst security engineers. The reason for this is three fold: 1) ideally, cyber-insurance increases Internet user safety because the insured increases self-defense as a rational response to the increase in insurance premium \cite{kmy1}\cite{kmy2}\cite{bs}\cite{yd}. This fact has also been mathematically proven by the authors in \cite{leb3}\cite{leb}, 2) in the IT industry, the mindset of `absolute protection' is slowly changing with the realization that absolute security is impossible and too expensive to even approach, while adequate security is good enough to enable normal functions - the rest of the risk that cannot be mitigated can be transferred to a third party \cite{kmy3}, and 3) cyber-insurance will lead to a market solution that will be aligned with economic incentives of cyber-insurers and users (individuals/organizations) - the cyber-insurers will earn profit from appropriately pricing premiums, whereas users will seek to hedge potential losses. In practice, users generally employ a simultaneous combination of retaining, mitigating, and insuring risks \cite{bs2}.

\subsection{Cyber-insurance and Information Asymmetry} 
Sufficient evidence exists in daily life (e.g., in the form of auto and health insurance) as well as in the academic literature (specifically focused on cyber-insurance\cite{kmy1}\cite{kmy2}\cite{leb3}\cite{leb}\cite{bs} that insurance-based solutions are useful approaches to pursue, i.e., as a complement to other security measures (e.g., anti-virus software). However, despite all promises, current cyber-insurance markets are \emph{non-competitive}, \emph{specialized}, and \emph{non-liquid}. The inability of cyber-insurance in becoming a common reality is due to a number of unresolved research challenges as well as practical considerations. The most prominent amongst them are \emph{information asymmetry} between the insurer and the insured, and the interdependent and correlated nature of cyber-risks \cite{rmb}\cite{rabohme}. Information asymmetry has a significant effect on most insurance environments, and is comprised of two components: (i) the inability of the insurer to distinguish between users of different (high and low risk) types, i.e., the \emph{adverse selection} problem, and (ii) users undertaking actions (i.e., reckless behavior) that affect loss probability \emph{after} the insurance contract is signed knowing that they would be insured, i.e., the \emph{moral hazard} problem. In the Internet, or as a matter of fact in any distributed communication network, some examples of information asymmetry that could arise due to (i) insurers lacking vital information regarding applications, software products installed by Internet users, and security maintenance habits, which correlate to the risk types of users, and (ii) users hiding information about their reckless behavioral intentions from their insurers, after they get insured, knowing that they would be compensated irrespective of their malicious behavior (e.g., accessing malicious websites, being careless with security settings, etc.,). This behavior by users affects the overall network security strength and might cause financial loss to cyber-insurers. 

\subsection{Our Research Contributions}
In this paper we model realistic, i.e., imperfect\footnote{A perfect insurance market is one where there is no information asymmetry between the cyber-insurer and the insured.}, single insurer (e.g., ISP or a government agency) cyber-insurance markets for distributed network environments and \emph{jointly} address the \emph{adverse selection }and \emph{moral hazard} problem in cyber-insurance. (See Section III). We design optimal cyber-insurance contracts under information asymmetry scenarios. Our design mechanisms are based on the \emph{Principal-Agent} (PA) model, which is built upon the theory of mechanism design in microeconomic theory \cite{mwg} (See Section IV). PA modeling is considered a powerful tool used in microeconomic theory to tackle situations of information non-transparency between economic entities \cite{mwg}. As part of our results in Section IV, we mathematically show
\begin{enumerate}
 \item Optimal cyber-insurance contracts induced by our mechanisms only provide partial coverage to the insureds, thereby ensuring greater self-defense efforts on the part of the latter to protect their computing systems, which in turn increase overall network security.
\item The level of insurance deductible charged per network user increases in a concave manner with increase in the topological degree of the user. 
\item A market for cyber-insurance \emph{can be made} to exist\footnote{ A situation of market equilibrium where both the insurers, as well as their clients are well-off with respect to their insurance contracts.} in the presence of monopolistic insurers under effective mechanism design, provided buying insurance is made mandatory for users. This result takes a step forward on the result in \cite{ssfw}, where the authors prove the non-existence of market equilibrium under the \emph{absence} of mechanism design. 
\end{enumerate}  

\section{Related Work}
In this section we briefly survey existing research work on cyber-insurance under the following two categories.
\subsection{Self-Defense Investments and Cyber-insurance}
The field of cyber-insurance in networked environments has been triggered by recent results on the amount of individual user self-defense investments in the presence of network externalities. The authors in \cite{gccr}\cite{jaw}\cite{leb5}\cite{leb4}\cite{mybm}\cite{oom} mathematically show that Internet users invest too little in self-defense mechanisms relative to the socially efficient level, due to the presence of network externalities. These works just highlight the role of positive externalities in preventing users from investing optimally in self-defense investments. Thus, the challenge to improving overall network security lies in incentivizing end-users to invest in sufficient amount of self-defense investments inspite of the positive externalities they experience from other users investing in the network. In response to the challenge, the works in \cite{leb5}\cite{leb4} modeled network externalities and showed that a tipping phenomenon is possible, i.e., in a situation of low level of self-defense, if a certain fraction of population decides to invest in self-defense mechanisms, it could trigger a large cascade of adoption in security features, thereby strengthening the overall Internet security. However, they did not state how the tipping phenomenon could be realized in practice. In a series of recent works \cite{leb3}\cite{leb}, Lelarge and Bolot have stated that under conditions of no \emph{information asymmetry} \cite{wik}\cite{hv} between the insurer and the insured, cyber-insurance \emph{incentivizes} Internet user investments in self-defense mechanisms, thereby paving the path to trigger a cascade of adoption. They also show that investments in both self-defense mechanisms and insurance schemes are quite inter-related in maintaining a socially efficient level of security on the Internet.
\subsection{Tackling Information Asymmetry}
Inspite of Lelarge and Bolot proposing the role of cyber-insurance for networked environments in incentivizing increasing user security investments, its common knowledge that the market for cyber-insurance has not blossomed with respect to its promised potential. Most recent works\cite{rabohme}\cite{ssfw} have attributed the underdeveloped market for cyber-insurance due to 1. \emph{interdependent security}, 2. \emph{correlated risk}, and 3. \emph{information asymmetries}. Thus, the need of the hour is to develop cyber-insurance solutions targeting these three issues, and identify other factors that might play an important role in promoting a developed cyber-insurance market. The works in \cite{hoffman}\cite{leb3}\cite{leb}\cite{rkk} touch upon the notion of information asymmetry and the effect it has on the insurance parameters, however none of the works \emph{explicitly} model information asymmetry. In relation to tackling information asymmetry, the authors in \cite{hoffman}\cite{leb3}\cite{ssfw} propose the concept of premium differentiation and fines to promote cyber-insurance. Another approach to resolving information asymmetry is via \emph{security auditing} \cite{amr}, where an auditing agency does an extensive introspection of the security behavior of an organization and passes on the information to an insurance agency, which in turn designs the optimal insurance contract based on the introspection report.  However, there are privacy concerns associated with this approach when it comes to handling non-organizational users, and might pose regulatory constraints upon the audit agency in the first place. 

\emph{Based on existing works it is clear that tackling information asymmetry formally has been an unchartered territory in cyber-insurance research.} Improving upon existing related works,  we take a first step in this direction and propose a formal model to resolve the information asymmetry problem in distributed communication networks. Assuming that cyber-insurance is made mandatory \cite{pgp}, our model enables existence of cyber-insurance markets, i.e., the existence of market equilibria, under non-ideal insurance environments. To the best of our knowledge, this is the first model of its kind specific to Internet and distributed network environments. 
\section{Model}
We structure this section in two parts. In the first part we describe the network environment and user utility functions pertaining to any distributed communication network. In the second part we model information asymmetry in cyber-insurance. We use the terms `user', `Internet user', and `network user' interchangeably to denote users in any communication network having Internet access. We also interchangeably use the terms `user' , `client', and `insured'. 

\subsection{Network Structure} \label{sec-ns}
We consider a set $N = \{1, ......, n\}$ of $n$ Internet users, where the connections between them form a graph $G = (V,E)$, where $v_{ij} = 1$ (edge weight between nodes (users) $i$ and $j$) if the utility of user $i$ is affected by the security (self-defense) investment of user $j$, $i$ being not equal to $j$, and 0 otherwise. Let $N_{i}(v) = \{j|v_{ij} = 1\}$ denote the set of all the one hop neighbors of $i$, where $v\,\epsilon\,\{0,1\}^{n \times n}$ is a matrix of connections amongst nodes. 
We represent the degree of a node $i$ by $d_{i}$, where $d_{i}$ equals $|N_{i}(v)|$. 

\subsection{User Payoffs} \label{sec-payoff}

We model the utility/payoff to each user $i$ as $U_{i}$, which is a function of the security investments made by himself, his one hop neighbors, and his final wealth\footnote{The final wealth is the net user wealth resulting after getting covered (uncovered) by a cyber-insurance policy in case of a loss (no-loss). }. We assume that the cyber-insurer knows the utility function of its clients\footnote{Such knowledge can in practice be estimated via surveys.}, and designs contracts based on it, and the \emph{type} of adverse selection scenario (See next subsection.). Mathematically, $U_{i} = U_{i}(W_{i}, x_{i}, \overrightarrow x_{N_{i}(v)})$, where $\overrightarrow x_{N_{i}(v)}$ is the vector of security investments of the one hop neighbors of user $i$, and $W_{i}$ is the final wealth of user $i$. From the structure of user utility functions, we observe that two users having the same degree and final wealth will have the same utility function. We also model the concept of a positive externality as it influences user self-defense investment decisions. A positive externality to a user from its one hop neighbors results when the latter invest in security, thereby improving the individual security strength of the user. We represent the concept mathematically in the following manner: we say that a payoff function exhibits positive externalities if for each $U_{i}$ and for all $\overrightarrow x \ge \overrightarrow x', U_{i}(x_{i}, \overrightarrow x, W_{i}) \ge U_{i}(x_{i}, \overrightarrow x', W_{i})$, where $\overrightarrow x$ and $\overrightarrow x'$ are the vectors of security investments of one hop neighbors of user $i$ and $W_{i}$ is the final wealth of user $i$. 

In scenarios where the security strength of a user $i$ depends on the sum of investments of himself and other neighboring users, we mathematically formulate $i$'s utility/payoff function as follows:
\begin{equation}
U_{i}(W_{i}, x_{i}, \overrightarrow x_{N_{i}(v)}) = f\left(x_{i} + \lambda\sum_{j = 1}^{d_{i}}x_{j}, x_{i}, W_{i}\right),
\end{equation}
where $f(\cdot)$ is a non-decreasing function of $\overrightarrow x$, $x_{i}$, and $W_{i}$. $\lambda$ is a real scalar quantity which determines the magnitude of the positive externality experienced by user $i$ due to the security investments made by his one-hop neighbors.  
  
In this paper we assume the utility functions of Internet users to be of the \emph{strategic substitute} type exhibiting \emph{positive externalities}. We say that a utility/payoff function exhibits strategic substitutes or is \emph{submodular} if it exhibits the property of decreasing differences, i.e., $U_{i}(x_{i}, \overrightarrow x, W_{i}) - U_{i}(x'_{i}, \overrightarrow x, W_{i}) \le U_{i}(x_{i}, \overrightarrow x', W_{i}) - U_{i}(x'_{i}, \overrightarrow x', W_{i})$. The practical interpretation of a strategic substitute as applicable to this paper is that an increase in the security investments of a user's neighbors reduces the marginal utility of the user, thus de-incentivizing him from investing. This happens due to the positive externality a neighbor exerts on the user through his own investments.

\subsection{Modeling Information Asymmetry}
We assume two classes of users (insured users), one which has a high chance of facing risks and the other which has a low chance. We term these classes as `LC' and `HC' respectively. Let $\theta, (1 - \theta)$ be the proportion of users who run a high chance(low chance) of facing risk of size $r$ respectively. However, on grounds of adverse selection the insurer cannot observe the class of any user. We consider two cases relevant to adverse selection in the Internet: 1) the insurer and/or the insured user have no knowledge about which risk class the insured falls in\footnote{This situation may generally happen when the users do not provide truthful information to insurance agency questionnaires and both the insurer as well as the insured cannot estimate the value of correlated and interdependent risks posed to individual insureds.} \emph{(most pertinent w.r.t. the Internet and communication networks.)} and 2) the insurer has no knowledge of a user's risk class but the user acquires this knowledge (through third-party agencies)\footnote{The third party agencies could be private organizations who might observe intrusions into user security, however such steps have regulatory and neutrality issues and thus are debatable in terms of practical implementation. We consider this case in the paper for modeling completeness.} before/after signing the contract but before it invests in self-defense investments. We assume that each user in class $i\,\epsilon\,\{LC,HC\}$ invests an amount $x_{i}$ in self-defense mechanisms after signing an insurance contract, which reduces its probability $p_{i}$ of being affected by Internet threats. We list the following mathematical properties related to our risk facing probability function $p$, for users in classes $LC$ and $HC$.
\begin{itemize}
\item $p(x)$ is a twice continuously differentiable decreasing function with $0 > p'_{LC}(x) > p'_{HC}(x)$ and $p''_{i}(x_{i}) > 0$, i.e., investments by users in class LC are more effective in reducing the loss probability than equivalent investments by users in class HC.
\item $p_{HC}(x) > p_{LC}(x)$.
\item 1 $> p_{HC}(x) \ge p_{LC}(x) > 0,\, \forall x\,\epsilon\,[0, \infty)$.
\end{itemize}
We model moral hazard by assuming that the cyber-insurer cannot observe or have knowledge about the amount of investments made by the insured after signing the insurance contract. Regarding user investments, apart from the self-defense investments made by a user, we assume a certain minimum amount of base investments of value $binv$ made by an Internet user of class $i$ \emph{prior} to signing insurance contracts, without which no user can be insured. Thus $p_{i}(binv)$ is the highest chance of risk a user of class $i$ may face.

The insurance company accounts for adverse selection and moral hazard and designs an insurance contracts of the form $C_{i} = (z_{i},c_{i})$, for all users $i$ in class $j\,\epsilon \,\{LC,HC\}$, where $z_{i}$ is the premium and $c_{i}$ is the net coverage\footnote{By the term `net-coverage' we mean the total coverage minus the premium costs. Note that we do not include self-defense investments as part of initial wealth of a user, but include the costs for investing in self-defense in the utility function $U_{i}$ for each user $i$.} for user $i$. An Internet user adopts the insurance contract and invests in self-defense mechanisms to achieve maximum benefit. We measure the benefit of users of a particular risk class $i$ as a utility, which is expressed as a function of contract $C_{i}$ and self-defense investments $x_{i}$. We define the expected utility function for users in risk class $i$ and facing a risk of value $r$\footnote{We assume an uniform value of risk for expositional simplicity.} as an expected utility of final wealth, and it is expressed as 
\begin{equation}
EU_{i}(C_{i}, x_{i}) = A + B,
\end{equation}
where 
\[A =  p_{i}(x_{i})U_{i}(w_{0} - r + c_{i}, x_{i},\overrightarrow x)\]
and 
\[B = (1 - p_{i}(x_{i}))U_{i}(w_{0} - z_{i}, x_{i}, \overrightarrow x).\]
Here $w_{i0}$ is the initial wealth of user $i$ and $x_{i}$ is the amount of self-defense investment he makes and $U_{i}()$ is an increasing continuously differentiable function ($U_{i}'(x_{i}) > 0, U_{i}''(x_{i}) < 0$) that denotes the utility of wealth. Differentiating Equation (1) w.r.t. $x_{i}$, we get the first order condition as
\begin{equation}
-p'_{i}(x_{i})[U_{i}(w_{i0} - z_{k}, x_{i}, \overrightarrow x) - U_{i}(w_{i0} - r + c_{k}, x_{i},\overrightarrow x)] = 0
\end{equation}
The first order condition generates the optimal self-defense investment, $x_{i}^{opt}$, for user $i$ that \emph{maximizes} his expected utility of final wealth. 

In the following sections we analyze optimal cyber-insurance contracts under the presence of moral hazard when 1) neither the insurer nor the insured has any information regarding the risk class of a user, 2) the insurer does not have information regarding user class but the insured acquires information after signing the contract but before making self-defense investments, and (3)  the insurer does not have information regarding user class but the insured acquires information before signing the contract. 

\section{Mechanisms For Alleviating Information Asymmetry}
In this section, we design three mechanisms to alleviate information asymmetry in cyber-insurance related to three different adverse selection scenarios mentioned in the previous section. For each mechanism, the outcome are the parameters of an optimal cyber-insurance contract, i.e., the \emph{coverage} and the \emph{premium.} 

\subsection{Neither the Insurer Nor the Insured Has Information} \label{sec-noinfo}
An Internet user $i$ does not know his risk class and therefore he maximizes his expected utility of final wealth by considering his probability of loss equal to an expected probability value of $p_{i}^{\alpha}(x) = \theta p_{HC}(x) + (1 - \theta)p_{LC}(x)$ and solving Equation (3). $\alpha$ could be considered as the risk class that each user feels he is in, as he does not have perfect information about whether he is in class $LC$ or $HC$\footnote{One could view $\alpha$ as an expected risk class/type a user feels he is in given that he does not know his actual risk type.}. We assume here that the values of $p_{LC}(x)$ and $p_{HC}(x)$ are common knowledge to the insurer and the insured. The cyber-insurer on the other hand, maximizes his profits by offering an optimal contract $(C_{i}^{\alpha})^{opt} = ((z_{i}^{\alpha})^{opt}, (c_{i}^{\alpha})^{opt})$. The optimization problem related to an insurer's profit is given as 
\[argmax_{z_{i}^{\alpha}, c_{i}^{\alpha},\lambda_{i}^{\alpha},\rho_{i}^{\alpha},\rho_{i}^{0}}[1 - p_{i}^{\alpha}(x_{i}^{\alpha})z_{i}^{\alpha} - p_{i}^{\alpha}(x_{i}^{\alpha})c_{i}^{\alpha}]\]
subject to
\begin{equation}
EU_{i}^{\alpha}((C_{i}^{\alpha})^{opt}, (x_{i}^{\alpha})^{opt}) - EU_{i}^{\alpha}(0,x_{i0}) \ge 0,
\end{equation}
\begin{equation}
-p_{i}^{\alpha'}(x_{i}^{\alpha})[U_{i}(w_{i0} - z_{i}^{\alpha}, x_{i}^{\alpha}, \overrightarrow x^{\alpha}) - U_{i}(w_{i0} - r + c_{i}^{\alpha}, x_{i}^{\alpha}, \overrightarrow x^{\alpha})] = 0,
\end{equation}
\begin{equation}
-p_{i}^{\alpha'}(x_{i0})[U_{i}(w_{i0}, x_{i}^{\alpha}, \overrightarrow x^{\alpha}) - U_{i}(w_{i0} - r, x_{i}^{\alpha}, \overrightarrow x^{\alpha})] = 0,
\end{equation}
where $x_{i0}$ is the amount of self-defense investments by user $i$ when no insurance is purchased. $\lambda_{i}^{\alpha},\rho_{i}^{\alpha},\rho_{i}^{0}$ are the Lagrangian multipliers related to constraints (4), (5), and (6) respectively. Constraint (4) is the participation constraint \emph{(Individual Rationality)} stating that the expected utility of final wealth of a user is atleast as much with cyber-insurance as without cyber-insurance. Constraints (5) and (6) state that Internet users will invest in optimal self-defense investments so as to maximize their utility of final wealth, and this is in exact accordance to what the cyber-insurer wants (i.e., to avoid moral hazard). 

The optimization problem presented in this section\footnote{We also note that the optimization problems in the forthcoming sections are all examples of general principal-agent problems.} is an example of a general \emph{principal-agent} problem. The Internet users (agents) will act non-cooperatively as utility maximizers, whereas the principal's (cyber-insurer) problem is to design a mechanism that maximizes its utility by accounting for adverse selection and moral hazard on the client (agent) side. Thus, the situation represents a \emph{Bayesian game of incomplete information} \cite{ft}. According to Palfrey and Srivastava \cite{ps}, there exists an \emph{incentive-compatible direct revelation mechanism} \cite{ngnp} for the problem implementable in private value models, where users do what the insurer desires (i.e., invest optimally in self-defense investments), provided the constraints in the optimization problem bind, and the users do not use \emph{weakly dominated strategies} \cite{ft} in equilibrium.

We have the following lemma stating the result related to the solution of the optimization problem. \\ \\
\textbf{Lemma 1.} \emph{The optimal cyber-insurance contract under situations when neither the insurer nor the insured have perfect information on the risk type of the client, induces a partial coverage at fair premiums. In addition, a pooling equilibrium (optimal) contract results for both high and low risk users.} \\ \\
\emph{Proof Sketch:} On route to solving our optimization problem, we derive the Lagrangian \cite{bv} and first order conditions, and apply the Karush-Kuhn-Tucker (KKT) conditions. We omit the proof in the paper due to lack of space. Details of proof methodology can be found in \cite{ltr}. \\
\emph{Lemma Implications:} The solution to the optimization problem in the binding case tends to \emph{full insurance} coverage as the utility function tends to become increasingly risk averse, and \emph{partial insurance} coverage otherwise. It also generates a \emph{pooling equilibrium} contract\footnote{A pooling equilibrium is one where the cyber-insurer has the same policy for both the classes (high and low risk) of users and the contract is in equilibrium.}, which is unique and entails partial cyber-insurance coverage at fair premiums. \emph{Thus, we infer that a partial insurance coverage is optimal for the cyber-insurer to provide to its clients as it accounts for the uncertainty of user risk types.} Intuitively, a pooling equilibrium works as neither the insurer nor the insured has any information on user risk type and as a result the cyber-insurer is not at a disadvantage regarding gaining risk type information relative to the Internet users. \emph{The pooling equilibrium establishes the existence of a market for cyber-insurance.}

\subsection{Insurer Has No Information, Insured Gets Information After Signing Contract} \label{sec-info}
In this scenario, we assume that the insurer does not have information about the risk class of a user and it cannot observe the risk class if the user obtains information from any third party agency. Since, the cyber-insurer is the first mover, it will account for the fact that users will be incentivized to take the help of a third party. 

Let $EU_{i}^{\alpha}(C_{i},x_{i})$ be the expected utility of user $i$ in risk class $\alpha$ for a contract $C_{i}$, when he cannot observe the risk class he is in. Let $\theta EU_{HC}(C_{i}, x_{i}) + (1 - \theta)U_{HC}(C_{i}, x_{i})$ be the expected utility of the same user when he can get information about his risk class from a third party agency. Thus, we denote the value to user $i$ of gaining information about his risk type w.r.t.  contract $C_{i}$ as $VI(C_{i})$, and it is defined for all $\theta\,\epsilon\,[0,1]$ as
\begin{equation}
VI(C_{i}) = \theta EU_{HC}(C_{i}, x_{i}) + (1 - \theta)EU_{HC}(C_{i}, x_{i}) - EU_{i}^{\alpha}(C_{i},x_{i}).
\end{equation}
We emphasize that $VI(C_{i})$ is zero if there is only type of risk class in the market. Now let $x_{ij}$ be the solution to Equation (3), for user $i$ being in risk class $j$ having contract $C_{i}$. Since $p'_{LC}(\cdot) < p'_{\alpha}(\cdot) < p'_{HC}(\cdot)$, for contract $C_{i}$, we have $x_{i}^{LC} > x_{i}^{\alpha } > x_{i}^{HC}$. Thus, $VI(C_{i}) > 0$ due to the following relationship
\begin{equation}
EU_{i}(C_{i}, x_{ij}) > EU_{i}(C_{i},x_{i}^{\alpha}),\,j\,\epsilon\,\{LC,HC\}.
\end{equation}
The cyber-insurer maximizes its profits by offering an optimal contract $C_{i}^{opt} = (z_{i}^{opt}, c_{i}^{opt})$. The optimization problem related to an insurer's profit is given as
\[argmax_{z_{i}, c_{i},\lambda_{i}^{j},\rho_{i}^{j},\rho_{i0}^{j}}\sum_{j = LC,HC}[1 - p_{i}^{j}(x_{i})z_{i} - p_{i}^{j}(x_{i})c_{i}]\]
subject to
\begin{equation}
EU_{i}^{j}(C_{i}^{opt}, x_{i}^{opt}) - EU_{i}^{j}(0,x_{i0}) \ge 0,\, j\,\epsilon\,\{ LC, HC\},
\end{equation}
\begin{equation}
-p_{i}^{j'}(x_{i})[U_{i}(w_{i0} - z_{i}, x_{i}, \overrightarrow x) - U_{i}(w_{i0} - r + c_{i}, x_{i}, \overrightarrow x)] = 0,\, \forall j,
\end{equation}
\begin{equation}
-p_{i}^{\alpha'}(x_{i0}^{j})[U_{i}(w_{i0}, x_{i}, \overrightarrow x) - U_{i}(w_{i0} - r, x_{i}, \overrightarrow x)] = 0,\, j\,\epsilon\,\{LC, HC\},
\end{equation}
where $x_{i0}$ is the amount of self-defense investments when no insurance is purchased by user $i$. $\lambda_{i}^{j},\rho_{i}^{j},\rho_{i0}^{j}$ are the Lagrangian multipliers related to constraints (9), (10), and (11) respectively. Constraint (11) is the participation constraint \emph{(Individual Rationality)} stating that the expected utility of final wealth of a user is atleast as much with cyber-insurance as without cyber-insurance. Constraints (10) and (11) state that Internet users will invest in optimal self-defense investments so as to maximize their utility of final wealth (moral hazard constraints).
We have the following lemma stating the result related to the solution of the optimization problem. The proof of the lemma follows from a similar proof sketch as that for Lemma 1. \\ \\
\textbf{Lemma 2.} \emph{The optimal cyber-insurance contract for each user $i$ induces a full coverage at fair premiums when $VI(C_{i}) = 0$ and induces partial coverage at fair premiums when $VI(C_{i}) > 0$. In addition, a pooling equilibrium (optimal) contract results for both high and low risk users.} \\ \\
\emph{Lemma Implications:} The solution to the optimization problem in the binding case results in \emph{full insurance} coverage if $VI(C_{i}) = 0$ and \emph{partial insurance} coverage if $VI(C_{i}) > 0$. If $VI(C_{k}) > 0$, which is most likely the case, a user would prefer to have information on its risk class and accept contract $C_{i}^{opt}$ rather than accept contract $(C_{i}^{\alpha})^{opt}$ (based on utility comparisons). Our optimization problem also generates a \emph{pooling equilibrium} contract, which is unique, and entails partial coverage at fair premiums. \emph{Thus, we infer that the cyber-insurer finds its optimal to provide partial insurance coverage to its clients as it accounts for uncertainty of user risk types.} Intuitively, a pooling equilibrium works as neither the insurer nor the insured has any information on user risk type \emph{before} the user signs the contract, and as a result the cyber-insurer is not at a disadvantage with respect to gaining information on risk type relative to Internet users.

\subsection{Insurer Has No Information, Insured Obtains Information Prior to Signing Contract} \label{sec-info1}
In this scenario, we assume that the insurer does not have information about the risk class of a user and it cannot observe the risk class if the user obtains information from any third party agency \emph{prior} to signing the insurance contract. However, in this scenario a user that knows his risk type is at a significant advantage. Since, the cyber-insurer is the first mover, he will account for the fact that users will be incentivized to take the help of a third party. We consider the case where the user may acquire information about his risk type prior to signing the insurance contract, and based on the information he decides on the contracts and in turn his self-defense investments.
We note here that users who remain uninformed will choose contract $C_{i}^{LC}$ as it is beneficial for the users to imitate the the low risk type users than be of the `expected' type.

We denote the value of gaining information to a user $i$ as $ VI = VI(C_{i}^{LC}, VI_{i}^{HC})$ and it is defined for all $\theta\,\epsilon\,[0,1]$ as
\begin{equation}
VI = EU_{i}^{HC}(C_{i}^{HC}, x_{i}^{HC}) + (1 - \theta)EU_{i}^{LC}(C_{i}^{LC}, x_{i}^{LC}) - EU_{i}^{\alpha}(C_{i}^{LC},x_{i}^{LC}).
\end{equation}
The cyber-insurer maximizes its profits by offering an optimal contract $C_{i}^{opt} = (z_{i}^{opt}, c_{i}^{opt})$. The optimization problem related to an insurer's profit is given as
\[argmax_{z_{i}, c_{i},\lambda_{i}^{j},\gamma_{i}^{k},\rho_{i}^{j},\rho_{i0}^{j}}\sum_{j = LC,HC}[1 - p_{i}^{j}(x_{i})z_{i} - p_{i}(x_{i})c_{i}]\]
subject to
\begin{equation}
EU_{i}^{j}(C_{i}^{opt}, x_{i}^{opt}) - EU_{i}^{j}(0,x_{i0}) \ge 0,\, j\,\epsilon\,\{LC, HC\},
\end{equation}
\begin{equation}
EU_{i}^{j}(C_{i}^{opt}, x_{i}^{opt}) - EU_{i}^{j}(C_{i}^{k},x_{i}^{k}) \ge 0,\, j,k\,\epsilon\,\{LC, HC\},
\end{equation}
\begin{equation}
-p_{i}^{j'}(x_{i})[U_{i}(w_{i0} - z_{i}, x_{i}, \overrightarrow x) - U_{i}(w_{i0} - r + c_{i}, x_{i}, \overrightarrow x)] = 0,\, j\,\epsilon\,\{LC, HC\},
\end{equation}
\begin{equation}
-p_{i}^{j'}(x_{k})[U_{i}(w_{i0} - z_{k}, x_{i}, \overrightarrow x) - U_{i}(w_{i0} - r + c_{k}, x_{i}, \overrightarrow x)] = 0,\, j,k\,\epsilon\,\{LC, HC\},
\end{equation}
\begin{equation}
-p_{i}^{\alpha'}(x_{i0})[U_{i}(w_{i0}) - U_{i}(w_{i0} - r)] = 0,\, j\,\epsilon\,\{LC, HC\},
\end{equation}
where $x_{i0}$ is the amount of self-defense investments when no insurance is purchased. $\lambda_{i}^{j},\gamma_{i}^{k},\rho_{i}^{j},\rho_{i0}^{j}$ are the Lagrangian multipliers related to constraints 13-17 respectively. Constraint (13) is the participation constraint stating that the expected utility of final wealth of a user is atleast as much with cyber-insurance as without cyber-insurance \emph{(Individual Rationality)}. Constraint (14) is the \emph{incentive compatibility} constraint, which states that users prefer to accept contracts that are designed to appeal to their types. Constraints (15), (16), and (17) state that Internet users will invest in optimal self-defense investments so as to maximize their utility of final wealth. We have the following lemma stating the result related to the solution of the optimization problem. The proof of the lemma follows from a similar proof sketch as that for Lemma 1. \\ \\
\textbf{Lemma 3.} \emph{The optimal cyber-insurance contract for each user $i$ induces a partial coverage at fair premiums. In addition, a separating equilibrium (optimal) contract results for both high and low risk users.} \\ \\
\emph{Lemma Implications:} Our optimization problem generates a \emph{separating equilibrium} contract\footnote{A separating equilibrium is one where the cyber-insurer has different insurance contracts for both the classes (high and low risk) of users and the contract is in equilibrium.}, which is unique and entails partial cyber-insurance coverage at fair premiums. Intuitively, a separating equilibrium works as the cyber-insurer is aware of the fact that Internet users have risk type information before they lay down the contracts and thus plans different contracts for different types. In terms of optimal contracts and cyber-insurer profits, the insurer is worse off than in the no-information case because in the latter case, the insurer extracts all user surplus, whereas in the former case, it extracts full surplus from the low risk type users but only extracts partial surplus from high risk type users. The separating equilibrium establishes the existence of a market for cyber-insurance. We now have the following theorem whose proof follows from lemmas 1, 2, and 3, and the fact that insurance purchase needs to  be made mandatory for users \cite{pgp}. \\ \\
\textbf{Theorem 1.} \emph{A market for cyber-insurance can be made to exist amongst risk-averse Internet users when (i) effective mechanism design is used to alleviate information asymmetry scenarios and (ii) it is mandatory for users to buy cyber-insurance.} \\ \\
\textbf{Comment:} We note that in the optimization problems stated above, the output is \emph{only} the optimal premium and coverage Through partial coverage we shift \emph{additional} liability to users to increase their investment amounts (atop the optimal efforts enforced in the problem constraints), thereby leading to increased overall security.  
\subsection{Effect of Topology on Contract Parameters}
In this section, we briefly present and analyze results related to the effect of user degrees on their cyber-insurance contract parameters. We have the following lemma relating user degrees with cyber-insurance coverage. We omit the proof in the paper due to lack of space. However, the proof concept (sketch) relies simply on evaluating the first derivative and second derivative of the deductible expression in the contract. \\ \\
\textbf{Lemma 4.} \emph{The level of deductible (coverage) for each Internet user $i$ on a risk of size $r$ increases (decreases) in a concave (convex) manner with the degree of the user, i.e., $\frac{dc_{i}}{dd_{i}} \le 0,\,\forall i$ and $\frac{d^{2}c_{i}}{dd_{i}^{2}} \ge 0,\,\forall i$, under every adverse selection scenario. }\\
\emph{Lemma Implications:} The intuition for Lemma 4 holding true is the fact that with increase in user degrees one gets well connected with his neighbors and invests less in self-defense investments but gains greater expected utility than his lesser connected counterparts \cite{ph}. This leads to a free riding phenomenon. Optimal cyber-insurance contracts for users derived in this paper accounts for this fact and introduces a control in terms of imposing higher deductibles (lesser coverage) to well connected users, hence incentivizing them to invest more in self-defense investments. 

\section{Conclusion}
In this paper we used the principal-agent model in microeconomic theory to address the information asymmetry problem in cyber-insurance and proposed mechanisms to alleviate the problem. The optimal contracts derived from our theory accounts for the topological location of each user in a communication network, enforce Internet users to take more responsibility in protecting their computing systems, and incentivizes them to increasingly invest in self-defense mechanisms. This in turn increases the overall network security. Through our mechanisms we also showed the existence of single-cyberinsurer insurance markets for Internet security. As part of future work, we plan to target multi-insurer cyber-insurance markets.

\bibliography{alluvion12}
\bibliographystyle{plain}


\end{document}